\documentclass[slac_one]{revtex4}
\usepackage{graphicx}
\usepackage{fancyhdr}
\begin{document}

\title{Ultra High Energy Cosmic Rays}

%

\author{Todor Stanev}
\affiliation{Bartol Research Institute, University of Delaware, Newark, DE 19716}

\begin{abstract}
  We discuss theoretical issues and experimental data that brought the 
 ultra high energy cosmic rays in the list of Nature's greatest puzzles.
 After many years of research we still do not know how astrophysical 
 acceleration processes can reach energies exceeding 10$^{11}$ GeV.
 The main alternative {\em top-down} mechanism postulates the existence
 of super massive $X$-particles that create a particle spectrum extending
 down to the observed energy through their decay channels.
 The propagation of nuclei and photons from their sources to
 us adds to the puzzle as all particles of these energies interact with
 the ambient photons, mostly of the microwave background. 
 We also describe briefly the main observational results and give some
 information on the new experiments that are being built and designed 
 now.
\end{abstract}

\maketitle

\thispagestyle{fancy}


\section{INTRODUCTION}
  More than forty years ago, in 1963, John Linsley~\cite{Linsley63} 
 published an article about the detection of a cosmic ray of energy 
 10$^{20}$ eV. The article did not go unnoticed, neither it provoked
 many comments. The few physicists that were interested in high energy
 cosmic rays were then convinced that the cosmic ray energy spectrum
 can continue forever. The fact that cosmic rays may have energies
 exceeding 10$^6$ GeV (10$^{15}$ eV) was established in the late 
 thirties by Pierre Auger and his collaborators. Showers of higher
 and higher energy were detected in the mean time - seeing a
 10$^{20}$ eV shower seem to be only a matter of time and exposure. 
 Already in the fifties there was a discussion about the origin
 of such ultra high energy cosmic rays (UHECR) and Cocconi~\cite{Cocconi56}
 reached the conclusion that they must be of extragalactic origin
 since the galactic cosmic rays are not strong enough contain such
 particles.

  How exclusive this event is became obvious three years later, after
 the discovery of the microwave background. Almost simultaneously
 Greisen in US~\cite{G} and Zatsepin\&Kuzmin~\cite{ZK} in the USSR
 published papers 
 discussing the propagation of ultra high energy particles in 
 extragalactic space. They calculated the energy loss distance
 of nucleons interacting in the microwave background and reached
 the conclusion that it is shorter than the distances between
 powerful galaxies. The cosmic ray spectrum should thus have an
 end around energy of 5$\times$10$^{19}$ eV. This effect is now
 known as the GZK cutoff. 

  The experimental statistics of such event grew with years, although
 not very fast. The flux of UHECR of energy above 10$^{20}$ eV is 
 estimated to 0.5 to 1 event per square kilometer per century per
 steradian. Even big detectors of area tens of km$^2$ would only
 detect  few events for ten years of work. The topic became one of
 common interest during the last decade of the last century when
 ideas appeared for construction of detectors with effective areas in
 thousand of km$^2$. Such detectors would detect tens of events per year
 and finally solve all mysteries surrounding UHECR, which I will attempt
 to convey to you.

\begin{figure}[thb]
\includegraphics[width=100mm]{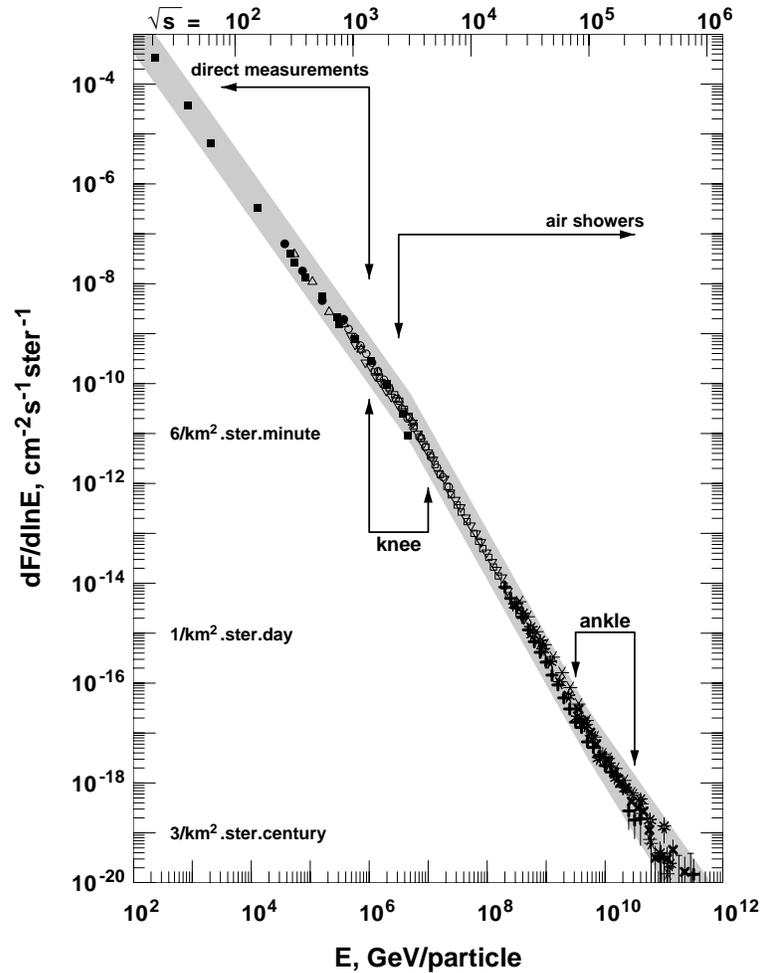}
\caption{Energy spectrum of the cosmic ray nuclei.}
\label{ts:fig1}
\end{figure}

 Cosmic rays are usually defined as charged nuclei that originate
 outside the solar system. They come on a featureless, power law like,
 $F(E) = K \times E^{-\alpha}$, spectrum that extends beyond
 10$^{11}$ GeV per particle, as shown above 100 GeV in Fig.~\ref{ts:fig1}.
 There are only two distinct features in the whole spectrum.
 At energy above 10$^6$ GeV the power law index $\alpha$ steepens
 from 2.7 to about 3.1. This is called {\em the knee} of the cosmic
 ray spectrum. At energy above 10$^9$ GeV the spectrum flattens again
 at {\em the ankle}. Both energy ranges, which are still not very well
 measured are indicated in the figure.

 The common wisdom is that cosmic rays below the knee are accelerated 
 at galactic supernovae remnants. Cosmic rays above the knee are also
 thought to be of galactic origin, although there is no clue of their
 acceleration sites. Cosmic rays above the ankle are thought to be
 extragalactic. When the total energy range is included the flux of
 cosmic rays varies by more than 30 orders of magnitude. Note that
 Fig.~\ref{ts:fig1} shows $E dF/dE$ and its slope is $\gamma = \alpha - 1$.
 The numbers of the left hand side of the figure shows the flux of
 cosmic rays in `natural' units. It is obvious that direct experiments
 for cosmic ray detection, that are flown in satellites or balloons,
 can not carry forever because of statistical limitations. Some data 
 from direct measurements are plotted with full symbols. All other
 symbols come from air shower measurements.

\subsection{Air Shower Detection Methods.}

 Cosmic rays of energy above 10$^{14}$ eV are detected by the showers
 they generate in the atmosphere. The atmosphere contains more than
 ten interaction lengths even in vertical direction and is much deeper
 for particles that enter it under higher zenith angles. It is thus
 a deep calorimeter in which the showers develop, reach their maximum,
 and then start being absorbed. There are generally two types of air
 shower detectors: air shower arrays and optical detectors.
 Air shower arrays consist of a large number of particle detectors
 that cover large area. The shower triggers the array by coincidental
 hits in many detectors. The most numerous particles in an air shower
 are electrons, positrons and photons. The shower also contains 
 muons, that are about 10\% of all shower particles, and hadrons
 (see the lecture of G.~Schatz for more information).

 The direction of the primary particle can be reconstructed quite well
 from the timing of the different hits, but the shower energy requires
 extensive Monte Carlo work with hadronic interaction models that are
 extended orders of magnitude above the accelerator energy range.
 The type of the primary particles can only be studied in statistically
 big samples because of the fluctuations in the individual shower 
 development. Even then it is strongly affected by the differences
 in the hadronic interaction models.

 The optical method uses the fact that part of the particle ionization
 loss is in the form of visible light. All charged particles emit
 in air Cherenkov light in a narrow cone around their direction. 
 In addition to that charged particles excite Nitrogen atoms in the
 atmosphere, which emit fluorescence light. The output is not large,
 about 4 photons per meter, but the number of shower particles
 in UHECR showers is very large, and the shower can be seen from 
 distances exceeding 30 km. The fluorescence detection is very suitable
 for UHECR showers because the light is emitted isotropically and
 can be detected independently of the shower direction.
 Since optical detectors follow the shower track, the direction of 
 the primary cosmic ray is also relatively easy. The energy of the 
 primary particles is deduced from the total number of particles 
 in the shower development, or from the number of particles at
 shower maximum. The rough number is that every particle at maximum
 carries about 1.5 GeV of primary energy. The mass of the primary
 cosmic ray nucleus is studied by the depth of shower maximum $X_{max}$,
 which is proportional to the logarithm of the primary energy per nucleon
 while the total energy depends on the number of particles at maximum.
   
 \subsection{The Highest Energy Cosmic Ray Event}

\begin{figure}[thb]
\includegraphics[height=70mm]{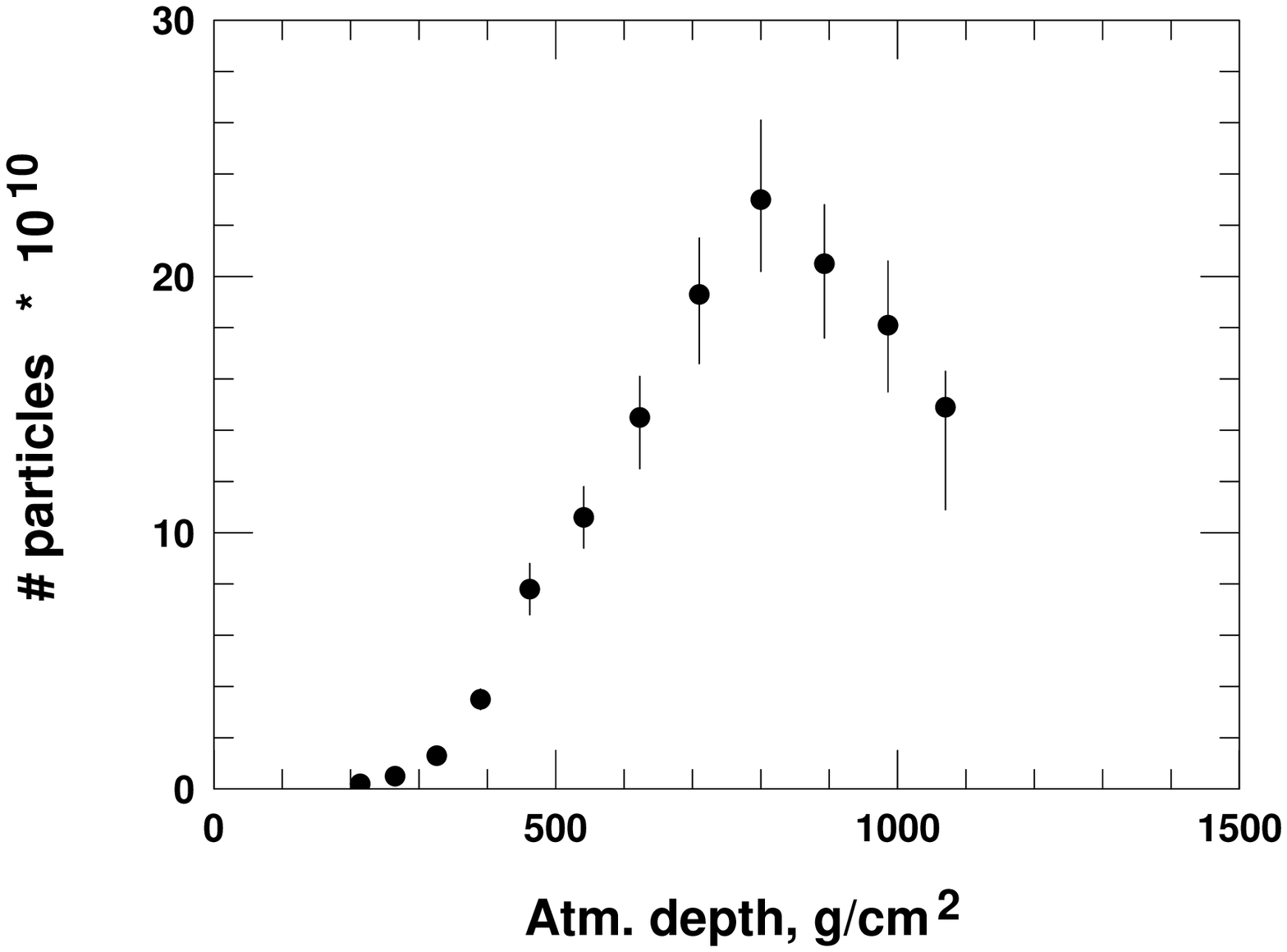}
\includegraphics[height=70mm]{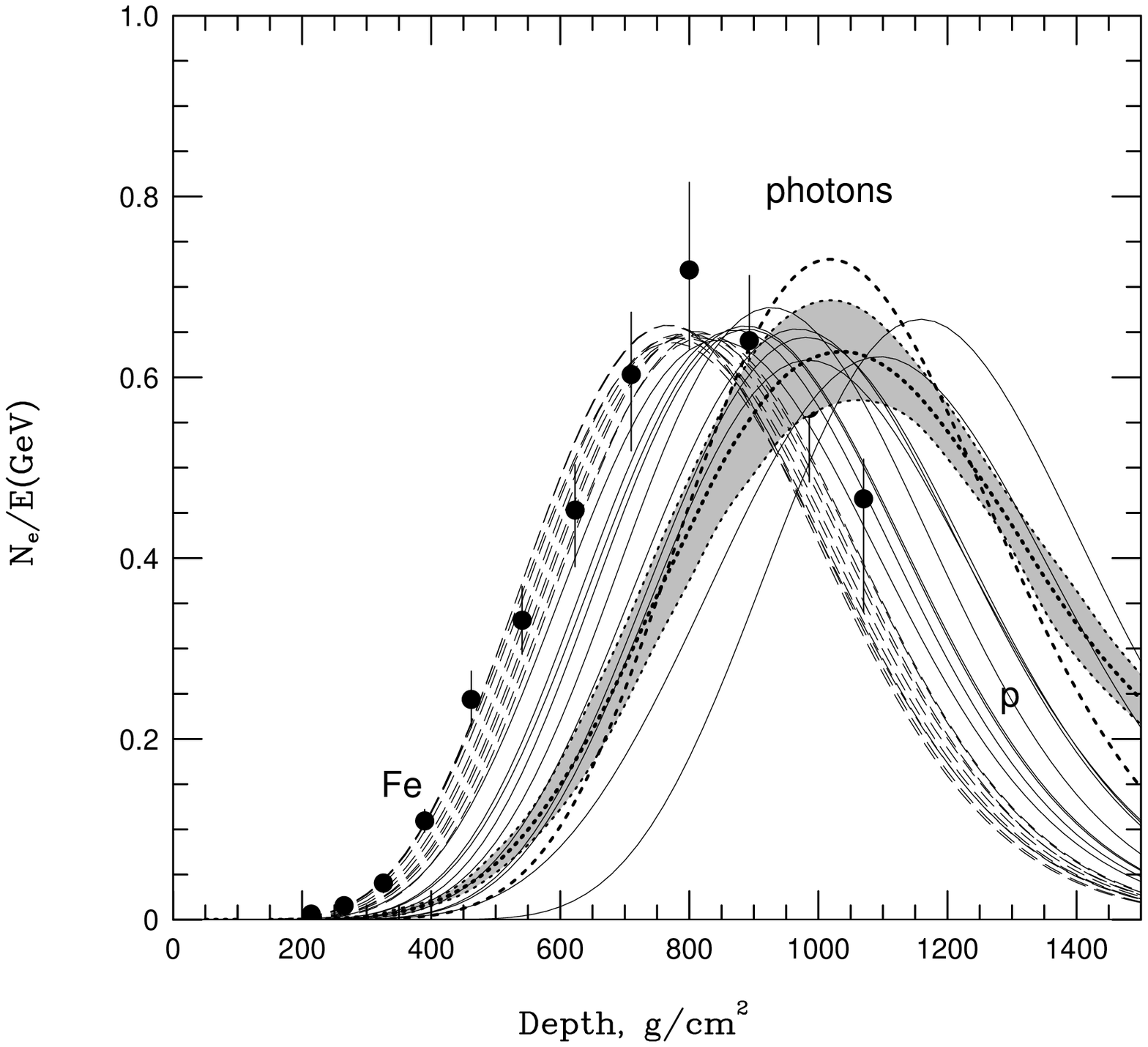}
\caption{Left hand panel: shower profile of the highest energy
 cosmic ray shower detected by the Fly's Eye. Right hand panel:
 comparison of the detected shower profile to Monte Carlo 
 calculations of shower initiated by protons (solid lines),
 iron nuclei (dashed lines), and gamma rays (shaded area, dots)
 in different assumptions for the development of gamma ray showers.}
\label{ts:fig2}
\end{figure}

 The highest energy cosmic ray particle was detected by the Fly's Eye 
 experiment~\cite{Bird_high}. We will briefly describe this event
 to give the reader an idea about these giant air showers. 
 The energy of the shower is estimated to be
 3$\times$10$^{20}$ eV. This is an enormous macroscopic energy.
 10$^{20}$ eV is equivalent to 1.6$\times$10$^{8}$ erg,
 2.4$\times$10$^{34}$ Hz and the energy of 170 km/h tennis ball.
 Fly's Eye was the first air fluorescence experiment, located in
 the state of Utah, U.S.A. Fig.~\ref{ts:fig2} shows the shower 
 profile of this event as measured by the Fly's Eye in the left
 hand panel. Note that 
 the maximum of this shower contains more than 2$\times$10$^{11}$
 electrons and positrons. Both the integral of this shower profile
 and the number of particles at maximum give about the same 
 energy.

 The errors of the estimates come from the errors of the individual
 data points, but mostly from the uncertainty in the distance 
 between the detector and the shower axis. The minimum energy of about
 2$\times$10$^{20}$ eV was calculated in the assumption that the
 shower axis was much closer to the detector than the data analysis
 derived. 

 It is very difficult to judge what type of the primary particle
 initiated the shower. A comparison of the detected shower
 profile with 10 showers initiated by protons and by primary iron
 nuclei is shown in the right hand panel of Fig.~\ref{ts:fig2}.
 It looks like the event is more consistent with the average event
 initiated by a light nucleus, but the fluctuations in the shower
 development make this notion very uncertain. On the other hand, if
 the same calculation were done with some of the different hadronic
 interaction models, it would be fully consistent with a proton 
 induced shower.

 As we shall see later, one of the crucial questions is if
 the shower was initiated by a primary nucleus, or by a primary
 $\gamma$-ray. Fig.~\ref{ts:fig2} attempts to answer this question.
 Gamma ray showers are usually no a subject to very big fluctuations,
 even after accounting for the LPM effect, which decreases the 
 pair production cross section at very high energy. The claim from
 this early analysis~\cite{HVSV} is that the detected shower profile is
 very different from the profiles of $\gamma$-ray showers in the 
 corresponding energy range, and thus the primary particle if 
 not a $\gamma$-ray.

\section{ORIGIN OF UHECR}

  The first problem with the ultra high energy cosmic rays is
 that it is very difficult to imagine what their origin is. 
 We have a standard theory for the acceleration of cosmic rays 
 of energy below the knee of the cosmic ray spectrum at 
 galactic supernova remnants. This suggestion was first made
 by Ginzburg\&Syrovatskii in 1960's on the basis of energetics.
 The estimate was that a small fraction (3-5\%) of the kinetic
 energy of galactic supernovae is sufficient to maintain
 the energy carried by the galactic cosmic rays. The acceleration 
 process was assumed to be stochastic, Fermi type, acceleration
 that was later replaced with the more efficient acceleration at
 astrophysical shocks. See R.~Blandford's lecture about 
 different acceleration models.

  This statement stands, but it is not applicable to all cosmic
 rays. Much more exact recent estimates and calculations show that
 the maximum energy achievable in acceleration on supernova
 remnant shocks is not much higher than 10$^5$ GeV. This excludes
 not only UHECR, but also the higher energy galactic cosmic rays,
 that require supernova remnants in special environments~\cite{VB}.

\begin{figure}[thb]
\includegraphics[height=80mm]{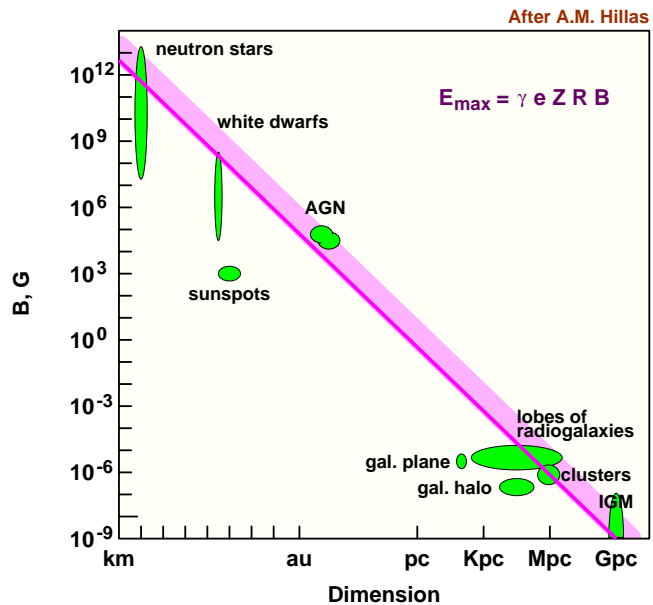}
\caption{Requirements for acceleration of charged nuclei at 
astrophysical objects as spelled out by A.M.~Hillas
(see text). 
}
\label{ts:fig3}
\end{figure}
 
  The reader should note that currently the acceleration of charged
 nuclei at supernova remnants is only a theoretical prediction.
 Supernova remnants have higher matter density than interstellar 
 space and one expects that the accelerated nuclei would interact 
 with the matter and generate high energy $\gamma$-rays. 
 Although many supernova remnants have been observed with TeV
 Cherenkov gamma ray telescopes, there is no proof that TeV 
 and higher energy $\gamma$-rays are generated in hadronic 
 interactions. We hope that the new generation of such telescopes:
 HESS, Magic, Kangaroo-3, and Veritas, will soon prove the prediction
 (see the lecture of W.~Hofmann).

  We should then turn to extragalactic objects for acceleration to
 energies exceeding 10$^{20}$ eV. The scale for such acceleration was
 set up by Hillas~\cite{Hillas83} from basic dimensional 
 arguments. The first requirement for acceleration of charged nuclei
 in any type of object is that the magnetic field of the object
 contains the accelerated nucleus within the object itself. 
 One can thus calculate a maximum theoretical acceleration energy,
 that does not include and efficiency factor, as 
 $$E_{max} \le \gamma e Z B R \; ,$$
 where $\gamma$ is the Lorentz factor of the shock matter, $Z$ is the
 charge of the nucleus, $B$ is the magnetic field value. and $R$ is
 the linear dimension of the object. 

  Figure~\ref{ts:fig3}, which is a redrawn version of the original 
 figure of Hillas, shows what are the requirements for acceleration 
 to more than 10$^{20}$ eV. The lower edge of the shaded area shows
 the minimum magnetic field value for acceleration of iron nuclei
 as a function of the dimension of the astrophysical object.
 The upper edge is for acceleration of protons. 

 There are very few objects that can, even before an account for
 efficiency, reach that energy: highly magnetized neutron stars,
 active galactic nuclei,
 lobes of giant radiogalaxies, and possibly Gpc size shocks from
 structure formation. Other potential acceleration sites, gamma
 ray bursts, are not included in the figure because of the time
 dependence of magnetic field and dimension. 

 \subsection{Possible Astrophysical Sources of UHECR}

 In this subsection we give a brief description of some of the models 
 for UHECR acceleration at specific astrophysical objects. For a more
 complete discussion one should consult a recent review paper on
 the astrophysical origin of UHECR~\cite{ta04}, that contains an
 exhaustive list of references to particular models.
 
\begin{itemize}
\item {\bf Pulsars:} Young magnetized neutron stars with surface
 magnetic fields of 10$^{13}$ Gauss can accelerate charged iron nuclei
 up to energies of 10$^{20}$ eV~\cite{beo00}. The acceleration process
 is magnetohydrodynamic, rather than stochastic as it is at 
 astrophysical shocks. The acceleration spectrum is very flat proportional
 to 1/$E$. It is possible that a large fraction of the observed UHECR
 are accelerated in our own Galaxy. There are also models for UHECR
 acceleration at magnetars, neutron stars with surface magnetic fields
 up to 10$^{15}$ Gauss.

\item {\bf Active Galactic Nuclei:} As acceleration site of UHECR
 jets~\cite{HZ97} of AGN have the advantage that acceleration on the jet frame 
 could have maximum energy smaller than these of the observed UHECR
 by 1/$\Gamma$, the Lorentz factor of the jet. The main problem with
 such models is most probably the adiabatic deceleration of the particles
 when the jet velocity starts slowing down.

\item {\bf Gamma Ray Bursts:} GRBs are obviously the most energetic
 processes that we know about. The jet Lorentz factors needed to model
 the GRB emission are of order 100 to 1000. These models became popular
 with the realization that the arrival directions of the two most energetic
 cosmic rays coincide with the error circles of two powerful GRB.
 Different theories put the acceleration site at the inner~\cite{Wax95}
 or the outer~\cite{Vietri} GRB shock. To explain the observed UHECRs 
 with GRBs one needs fairly high current GRB activity, while most of
 the GRB with determined redshifts are at $Z\; >$ 1.

\item {\bf Giant Radio Galaxies:} One of the first concrete
 model for UHECR acceleration is that of Rachen\&Biermann, that
 dealt with acceleration at FR II galaxies~\cite{rb93}. Cosmic
 rays are accelerated at the `red spots', the termination shocks 
 of the jets that extend at more than 100 Kpc. The magnetic fields
 inside the red spots seem to be sufficient for acceleration up to
 10$^{20}$ eV, and the fact that these shocks are already inside 
 the extragalactic space and there will be no adiabatic 
 deceleration. Possible cosmologically nearby objects include 
 Cen A (distance of 5 Mpc) and M87 in the Virgo cluster (distance
 of 18 Mpc).

\item {\bf Quiet Black holes:} These are very massive quiet black
 holes, remnants of quasars, as acceleration sites~\cite{bg99}.
 Such remnants could be located as close as 50 Mpc from our Galaxy.
 These objects are not active at radio frequencies, but, if massive
 enough, could do the job. Acceleration to 10$^{20}$ requires a mass
 of 10$^9$ M$_\odot$.

\item {\bf Colliding Galaxies:} These systems are attractive with the
 numerous shocks and magnetic fields of order 20 $\mu$G that have 
 been observed in them~\cite{CJC92}. The sizes of the colliding galaxies
 are very
 different and with the observed high fields may exceed the
 gyroradius of the accelerated cosmic ray.

\item {\bf Clusters of Galaxies:} Magnetic fields of order several
 $\mu$G have been observed at lengthscales of 500 Kpc. Acceleration
 to almost 10$^{20}$ eV is possible, but most of the lower energy
 cosmic rays will be contained in the cluster forever and only
 the highest energy particles will be able to escape~\cite{krj98}.
  
\item {\bf Gpc scale shocks from structure formation:} A combination
 of Gpc scales with 1 nG magnetic field satisfies the Hillas 
 criterion, however the acceleration at such shocks could be much
 too slow, and subject to large energy loss.
\end{itemize}  
 
\subsection{Top-Down Scenarios}

 Since it became obvious that the astrophysical acceleration up to
 10$^{20}$ eV and beyond is very difficult and unlikely, a large 
 number of particle physics scenarios were discussed as explanations
 of the origin of UHECR. To distinguish them from the acceleration
 ({\em bottom-up}) processes they were called {\em top-down}.
 The basic idea is that very massive (GUT scale) $X$ particles decay 
 and the resulting fragmentation process downgrades the energy 
 to generate the observed UHECR. Since the observed cosmic rays 
 have energies orders of magnitude lower than the $X$ particle mass,
 there are no problems with achieving the necessary energy scale.
 The energy content of UHECR is not very high, and the $X$ particles 
 do not have to be a large fraction of the dark matter.
 
 There are two distinct branches of such theories. One of them
 involves the emission of $X$ particles by topological defects.
 This type of models follows the early work of C.T.~Hill~\cite{hill83}
 who described the emission from annihilating monopole/antimonopole
 pair, which forms a metastable monopolonium. The emission of 
 massive $X$ particles is possible by superconducting cosmic string
 loops as well as from cusp evaporation in normal cosmic strings and
 from  intersecting cosmic strings. The $X$ particles then decay in
 quarks and leptons. The quarks hadronize in baryons and mesons,
 that decay themselves along their decay chains. The end result 
 is a number of nucleons, and much greater (about a factor of 30
 in different hadronization models) and approximately equal 
 number of $\gamma$-rays and neutrinos.   
   
 A monopole is about 40 heavier than a $X$ particle, so every monolonium
 can emit 80 of them. Using that number one can estimate the number
 of annihilations that can provide the measured UHECR flux, which turns
 out to be less than 1 per year per volume such as that of the Galaxy.
 Another possibility is the emission of $X$ particles from cosmic
 necklaces - a closed loop of cosmic string including monopoles. 
 This particular type of topological defect has been extensively
 studied~\cite{bv97}.
 
 The other option is that the $X$ particles themselves are remnants 
 of the early Universe. Their lifetime should be very long, 
 maybe longer than the age of the Universe~\cite{bkv97}. They could
 also be a significant part of the cold dark matter. Being superheavy,
 these particle would be gravitationally attracted to the Galaxy 
 and to the local supercluster, where their density could well exceed
 the average density in the Universe.  There is a large number of
 topological defect models which are extensively reviewed
 in Ref.~\cite{bs00}.
  
 There are two main differences between bottom-up and top-down models
 of UHECR origin. The astrophysical acceleration generates charged 
 nuclei, while the top-down models generate mostly neutrinos and
 $\gamma$-rays plus a relatively small number of protons. 
 The energy spectrum of the cosmic rays that are generated in the
 decay of $X$ particles is relatively flat, close to a power law
 spectrum of index $\alpha$=1.5. The standard acceleration energy
 spectrum has index equal to or exceeding 2.

\subsection{Hybrid Models}

 There also models that are hybrid, they include elements of both groups.
 The most successful of those is the Z-burst model~\cite{tjw99,fms99}.
  The idea is that somewhere in the Universe neutrinos of ultrahigh
 energy are generated. These neutrinos annihilate with cosmological
 neutrinos in our neighborhood and generate $Z_0$ bosons which decay
 and generate a local flux of nucleons, pions, photons and neutrinos.
 The resonant energy for $Z_0$ production is
 4$\times$10$^{21}$ eV/$m_\nu$(eV), where $m_\nu$ is the mass of the
 cosmological neutrinos. The higher the mass of the 
 cosmological neutrinos is, the lower the resonance energy requirement.
 In addition the cosmological neutrinos are gravitationally attracted 
 to concentrations of matter and their density increases in our
 cosmological neighborhood. If the neutrino masses are low, then the
 energy of the high energy neutrinos should increase.

\section{PROPAGATION OF UHECR}

 Particles of energy 10$^{20}$ eV can interact on almost any target.
 The most common, and better known, target is the microwave 
 background radiation (MBR). It fills the whole Universe and its number
 density of 400 cm$^{-3}$ is large. The interactions on the 
 radio and infrared backgrounds are also important. Let us have
 a look at the main processes that cause energy loss of nuclei
 and gamma rays.

\subsection{Energy Loss Processes}

 The main energy loss process for protons is the photoproduction on
 astrophysical photon fields $p\gamma \rightarrow p + n\pi$.
 The minimum center of mass energy
 for photoproduction is $\sqrt {s_{thr}} = m_p + m_{\pi^0}\; \sim$ 1.08 GeV.
 Since $s = m_p^2 + 2 (1 - \cos{\theta}) E_p \epsilon$
 (where $\theta$ is the angle between the two particles) one can
 estimate the proton threshold energy for photoproduction on the MBR
 (average energy $\epsilon\; =$ 6.3$\times$10$^{-4}$ eV).
 For $\cos{\theta}$ = 0 the proton threshold energy is
 $E_{thr}$ =  2.3$\times$10$^{20}$ eV. Because there are head to head
 collisions and because the tail of the MBR energy spectrum continues
 to higher energy, the intersection cross section is non zero above
 proton energy of 3$\times$10$^{19}$ eV.

\begin{figure}[thb]
\includegraphics[width=85mm]{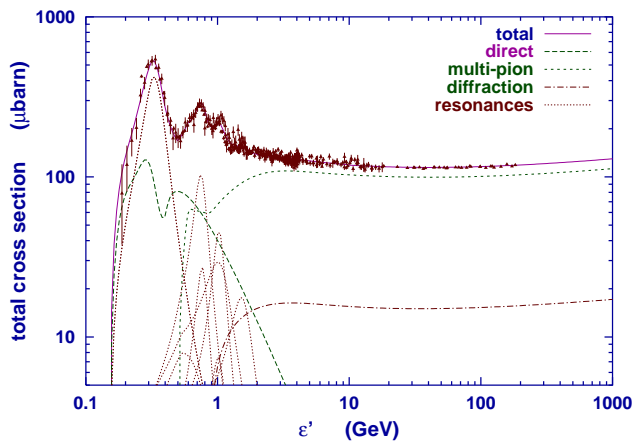}
\caption{Photoproduction cross section as a function of the
photon energy for stationary proton targets. 
}
\label{ts:fig4}
\end{figure}

 The photoproduction cross section is very well studied in accelerator
 experiments and is  known in detail. Figure~\ref{ts:fig3} shows the
 photoproduction cross section in the mirror system, as a function of
 the photon energy  for stationary protons, i.e. as it is measured in
 accelerators. At threshold the most important process is the 
 $\Delta^+$ production where the cross section reaches a peak exceeding
 500 $\mu$b. It is followed by a complicated range that includes the 
 higher mass resonances and comes down to about 100 $\mu$b. After that
 one observes an increase that makes the photoproduction cross
 section parallel to the $pp$ inelastic cross section. The neutron 
 photoproduction cross section is nearly identical.

 Another important parameter is the proton inelasticity $k_{inel}$,
 the fraction of its energy that a proton loses in one interaction.
 This quantity is energy dependent. At threshold protons lose
 about 18\% on their energy. With increase of the CM energy this
 fractional energy loss increases to reach asymptotically 50\%.

\begin{figure}[thb]
\includegraphics[width=85mm]{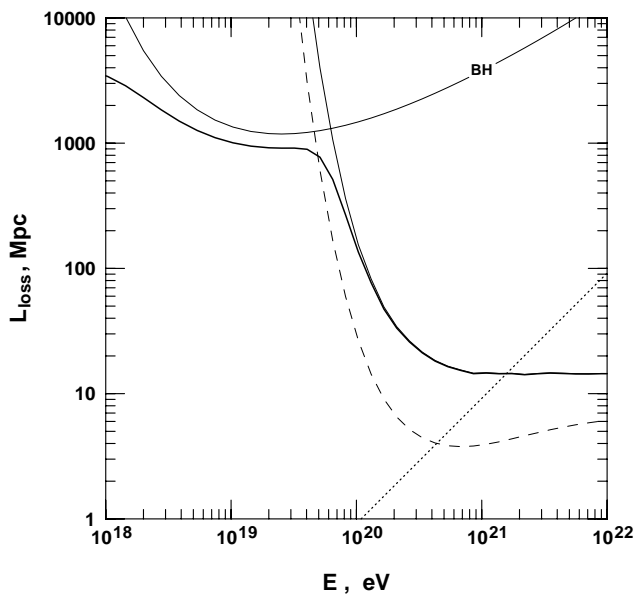}
\caption{Energy loss length of protons in interactions in MBR. 
}
\label{ts:fig5}
\end{figure}

 The proton pair production $p\gamma \rightarrow e^+e^-$ is the
 same process that all charged particles suffer in nuclear fields.
 The cross section is high, but the proton energy loss is of order
 $m_e/m_p$ $\simeq$ 4$\times$10$^{-4}$E. Figure~\ref{ts:fig5} shows the 
 energy loss length $L_{loss} = \lambda/k_{inel}$
 (the ratio of the interaction length to the inelasticity coefficient)
 of protons in interactions in the microwave background. 
 
 The dashed line shows the proton interaction length and one can see 
 the increase of $k_{inel}$ in the ratio of the interaction to energy
 loss length. The contribution of the pair production is shown with
 a thin line. The energy loss length never exceeds 4,000 Mpc,
 which is the adiabatic energy loss due to the expansion of the 
 Universe for $H_0$ = 75 km/s/Mpc. The dotted line shows the neutron
 decay length. Neutrons of energy less than about 3$\times$10$^{20}$ eV 
 always decay and higher energy neutrons only interact.

 Heavier nuclei lose energy to a different process - photodisintegration,
 loss of nucleons mostly at the giant dipole resonance~\cite{PSB76}.
 Since the  relevant energy in the nuclear frame is of order 20 MeV,
 the process starts at lower energy. The resulting nuclear fragment may
 not be stable. It then decays and speeds up the energy loss of the whole
 nucleus.
 Ultra high energy heavy nuclei, where the energy per nucleon is higher
 than photoproduction, have also loss on photoproduction. The energy loss 
 length for He nuclei in photodisintegration is as low as 10 Mpc at
 energy of 10$^{20}$ eV.
 Heavier nuclei reach that distance at higher total energy. 

\begin{figure}[thb]
\includegraphics[width=85mm]{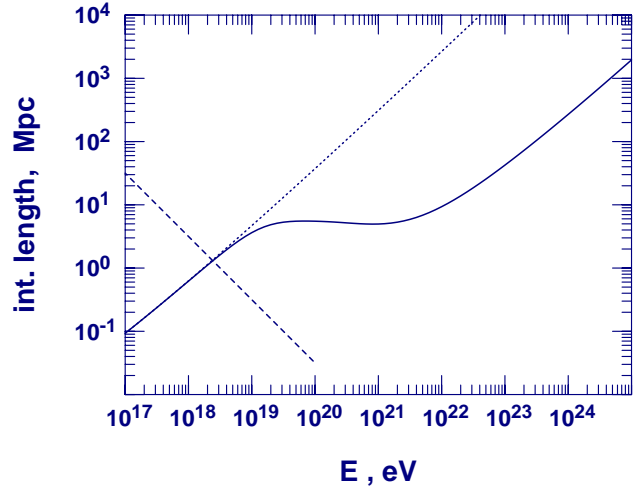}
\caption{Interaction length of \protect$\gamma$-rays in MBR (dots)
 and in MBR + radio background. The dashed line shows the synchrotron
 energy loss length of electrons in  1 nG magnetic field.}
\label{ts:fig6}
\end{figure}

 UHE gamma rays also interact on the microwave background. The main
 process is $\gamma \gamma \rightarrow e^+ e^-$. This is a resonant 
 process and for interactions in the MBR the minimum interaction 
 length is achieved at 10$^{15}$ eV. The interaction length in MBR
 decreases at higher $\gamma$-ray energy and would be about a 50 Mpc
 at 10$^{20}$ eV if not for the radio background. The radio background
 does exist but its number density is not well known. Figure~\ref{ts:fig6}
 shows the interaction length for this process in MBR (dots) and 
 in all photon fields. The energy range is wider than for protons
 because some top-down models can generate $\gamma$-rays of energy
 approaching $m_X$.

 The fate of the electrons produced in a $\gamma \gamma$ collision 
 depends on the strength of the magnetic fields in which UHE electrons
 lose energy very fast. the dashed line shows the electron energy loss
 length in 1 nG field. The photon energy is than quickly downgraded
 and the $\gamma \gamma$ interaction length becomes very close
 to the gamma ray energy loss length. In the case of very low magnetic
 fields (0.01 nG) the synchrotron energy loss is low (it is proportional
 to $E_e^2 B^2$) and then inverse Compton scattering (with a cross section
 very similar to this of $\gamma \gamma$) and cascading is possible.
 The energy loss length of the gamma rays would be
 higher in such a case.

\begin{figure}[thb]
\includegraphics[width=70mm]{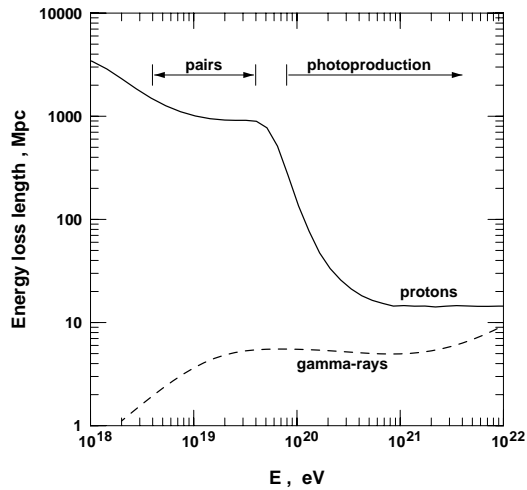}
\caption{Energy loss lengths of protons and gamma rays,
}
\label{ts:fig7}
\end{figure}

 In conclusion, the energy loss of protons, heavier nuclei and photons
 is high in propagation on cosmological scales. Figure~\ref{ts:fig7} 
 compares the proton energy loss length to that of gamma rays in the
 assumption of 1 nG magnetic fields.

 At energies below 10$^{20}$ eV the proton energy loss length is
 definitely longer than that of gamma rays. At energies above 
 5$\times$10$^{20}$ the difference is only a factor of 2, with
 very small energy dependence. Have in mind, though, that the
 flat part of the gamma ray energy loss length is due to 
 interactions in the radio background in the 1 MHz range, which can
 not be detected  at Earth and has to be modeled as a ratio
 to other astrophysical photon fields. 

 The general conclusion from this analysis of the energy loss of
 protons and gamma rays in their propagation through the Universe
 is these UHE particles can not survive at distances of more than
 few tens of Mpc and sources of the detected cosmic rays have to
 be located in our cosmological neighborhood. Every increase of
 the distance between the source and the observer would require
 and increase of the maximum energy at acceleration (or other
 production mechanism) and will increase significantly the 
 energy requirement to the UHECR sources.

\subsection{Modification of the Proton Spectrum in Propagation.
 Numerical Derivation\\ of the GZK Effect.}

\begin{figure}[thb]
\includegraphics[width=85mm]{read_prop1.ps}
\includegraphics[width=85mm]{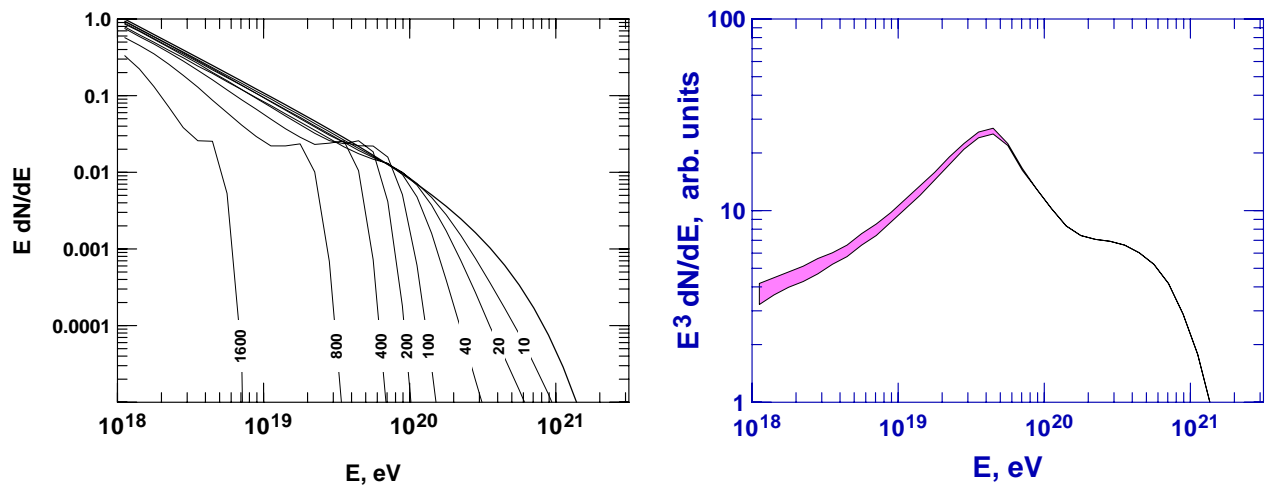}
\caption{ Left hand panel: Evolution of the cosmic ray spectrum 
 in propagation through different distances. Right hand panel:
 spectrum from homogeneous isotropic cosmic ray sources that
 inject spectra with \protect$\alpha$ = 2. Upper edge: cosmological
 evolution of UHECR sources with \protect$n$ = 4, lower one - \protect$n$ = 3.
}
\label{ts:fig8}
\end{figure}

 Figure~\ref{ts:fig8} shows in the left hand panel the evolution of
 the spectrum of protons 
 because of energy loss during propagation at different distances.
 The thick solid lines shows the spectrum injected in intergalactic
 space by the source, which in this exercise is 
$$ \frac{dN}{dE} \, = \, A \times E^{-2}/ \exp({E/3.16\times 10^{21}}
 {\rm eV}) \; .$$
 After propagation on 10 Mpc only some of the highest energy protons are
 missing. This trend continues with distance and at about 40 Mpc 
 another trend appears - the flux of protons of energy just below 10$^{20}$ eV
 is above the injected one. This is the beginning of the formation
 of a pile-up in the range where the photoproduction cross section
 starts decreasing. Higher energy particles that are downgraded in 
 this region lose energy less frequently and a pile-up is developed.

 The pile-up is better visible in the spectra of protons propagated at 
 larger distances. One should remark that the size of the pile-up 
 depends very strongly on the shape of the injected spectrum. If it
 had a spectral index of 3 instead of 2 the size of the pile-up 
 would have been barely visible as the number of high energy particles
 decreased by a factor of 10.

 When the propagation distance exceeds 1 Gpc there are no more 
 particles of energy above 10$^{19}$ eV. All these particles have lost
 energy in photoproduction, pair production and adiabatic losses.
 From there on most of the losses are adiabatic. 

 In order to obtain the proton spectrum created by homogeneously and
 isotropically distributed cosmic ray sources filling the whole
 Universe one has to integrate a set of such (propagated) spectra
 in redshift using the cosmological evolution of the cosmic ray
 sources, which is usually assumed to be the same as that of the star
 forming regions (SFR) $\eta(z) \; = \; \eta(0)(1 + z)^n$ with 
 $n$ = 3, or 4 up to the epoch of maximum activity $z_{max}$ and 
 then either constant or declining at higher redshift. High redshifts
 do not contribute anything to UHECR (1600 Mpc corresponds to 
 $z$ = 0.4 for $H_0$ = 75 km/s/Mpc). After accounting for the increased
 source activity the size of the pile-ups has a slight increase.
  
 The right hand panel shows the UHECR spectrum that comes from the
 integration of propagation spectra shown in the left hand panel 
 with different cosmological evolutions. Obviously the importance
 of the cosmological evolution is very small and totally disappears
 for very high energy. The differential spectrum is multiplied by 
 $E^3$ as is often done with experimental data to emphasize the
 spectral features. One can see the pile-up at 5$\times$10$^{19}$ eV
 after which the spectrum declines steeply. There is also a dip at about
 10$^{19}$ eV which is due to the energy loss on pair production.
 These features were first pointed at by Berezinsky \&
 Grigorieva~\cite{BerGri}. Such should be the energy spectrum of
 extragalactic protons under the assumptions of injection spectrum
 shape, cosmic ray luminosity (4.5$\times$10$^{44}$
 erg/Mpc$^3$/yr~\cite{Wax95}), cosmological evolution and isotropic
 distribution of the cosmic ray sources in the Universe.

\subsection{Modification of the Gamma Ray Spectra.}

\begin{figure}[thb]
\includegraphics[width=100mm]{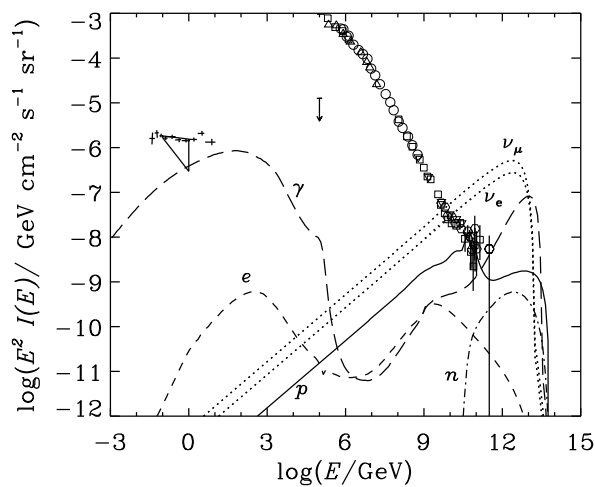}
\caption{ Evolution of the energy spectra of particles injected 
in a top-down scenario with \protect$X$ particle mass 
of 10\protect$^{14}$ GeV. The injection spectra a almost parallel 
to the shown neutrino spectra. 
}
\label{proth_sta}
\end{figure}

  Because of the strong influence of the radio background and of the
 cosmic magnetic fields the modification of the spectrum of gamma 
 rays in a top-down scenario is much more difficult to calculate
 exactly. There are, however, many general features that are common 
 in any of the calculations. Figure~\ref{proth_sta} shows the
 gamma ray spectrum emitted in a top-down scenario with $m_X$ =
 10$^{14}$ GeV~\cite{proth_sta}.
 The spectra of different particles from the $X$ 
 decay chain are indicated with different line types. The one, that
 we are now interested in, is for gamma rays. If not for energy
 loss the gamma ray spectrum would have been parallel to these
 of neutrinos.

 We shall start the discussion from the highest energy and follow 
 the energy dissipation in propagation.  
 The highest energy gamma rays have not suffered significant losses.
 At slightly lower energy, though, the $\gamma \gamma $ cross section
 grows and the energy loss increases. One can see the dip at about
 10$^{10 - 11}$ GeV which is caused by the radio background.
 All other interactions are on the MBR. The magnetic field is assumed
 sufficiently high that all electrons above 10$^9$ GeV immediately
 lose energy in synchrotron radiation. 
 The minimum ration of the gamma-ray to cosmic ray flux 
 is reached at about 10$^{15}$ eV, after 
 which there is some recovery. There is another absorption
 feature from interactions on the infrared background.
 The gamma ray peak in the GeV region consists mostly of synchrotron photons.
 Isotropic GeV gamma rays, that have been measured, can be used to
 restrict top-down models in some assumptions for the magnetic field
 strength. 
 
\subsection{UHECR Propagation and Extragalactic Magnetic Fields}

 The possible existence of non negligible extragalactic magnetic fields
 would significantly modify the propagation of the UHE cosmic rays
 independently of their nature and origin. There is little observational
 data on these fields. The best estimate of the average strength of
 these fields in the Universe is 10$^{-9}$ Gauss (1 nG)~\cite{Kronberg94}.
 On the other hand $\mu$G fields have been observed in clusters of 
 galaxies, and in a bridge between two parts of the Coma cluster.

 Since the measurements of fields of strength less than 1 $\mu$G are 
 very difficult, all current arguments are of strictly theoretical 
 type. Cosmological seed fields are not expected to be stronger than
 10$^{-17}$ G. Many people believe that the Universe is much too young
 to explain the existence of significant large scale fields by spreding
 out the fields of individual astrophysical objects.
 Others are
 impressed by the high fields observed in clusters of galaxies and
 suspect the existence of lower fields on larger scales.

 No one expects the extragalactic magnetic fields to be isotropic.
 They should be much higher than the average is the walls of high matter
 density and much lower in the voids. All possible UHECR sources should
 also be associated with high matter concentration.
  
 Even fields with nG strength would seriously affect the propagation of
 UHE cosmic rays. If UHECR are protons they would scatter of these fields.
 This scattering would lead to deviations from the source direction and 
 to an increase of the pathlength from the source to the observer.
 It would make the source directions less obvious and would create a
 magnetic horizon for extragalactic protons of energy below 10$^{19}$ eV
 as their propagation time from the source to the observer would 
 start exceeding Hubble time.

\begin{figure}[thb]
\includegraphics[height=60mm]{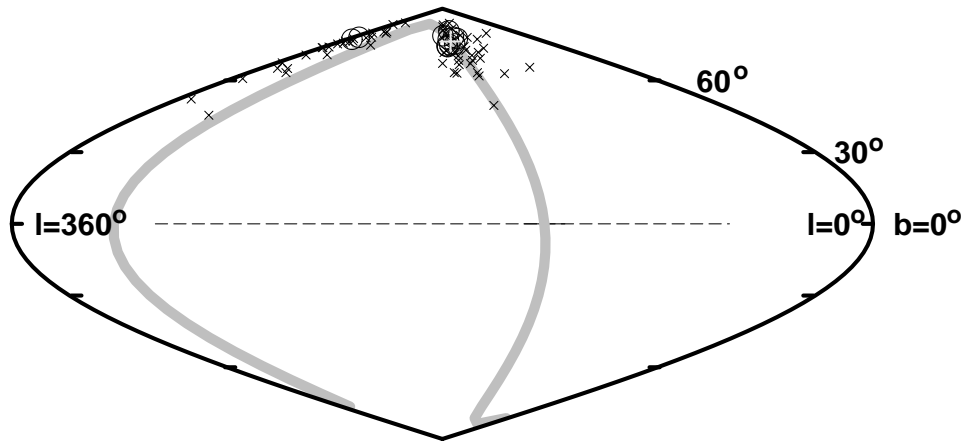}
\caption{The arrival direction in galactic coordinates of protons
 emitted by M87 (cross)
 if there were an organized 10 nG field along the Supergalactic plane
 (shown with wide gray curve). Particles above 10\protect$^{20}$ eV
 are shown with circles. The Galactic plane is indicated with a dashed
 line. The map is centered at the Galactic anticenter.
}
\label{cipanp}
\end{figure}

 If regular magnetic fields of strength exceeding 10 nG were present on
 10 Mpc coherence length they would lead to significant biases in the 
 propagated spectra~\cite{SSE03} in a function of the relative geometry
 of the field, source and observer. Particles of very high energy would
 gyrate around the magnetic field lines and thus appear coming from 
 a wide range of directions. Figure~\ref{cipanp} shows an example for
 protons accelerated at the powerful AGN M87 in case is connected with
 our galaxy by 3 Mpc wide cylinder of 10 nG magnetic field. The distance
 to M87 is 18 Mpc. If such fields indeed exist, one should be able to
 recognize the UHECR sources only on the basis of high statistics. 

 The strongest possible magnetic field effect on UHE cosmic ray nuclei
 was suggested by Biermann~\cite{PLB}. He used the observational fact
 that in all observed spiral galaxies the magnetic field is directed
 inward along the spiral arms and applied Parker's solar field
 model to obtain a model of the galactic wind.
 Galactic magnetic fields are frozen in the galactic wind.
 Protons of energy 10$^{20}$ eV can penetrate the resulting 
 magnetic bottle only from the galactic North and the most likely 
 origin of the detected UHECR is then in the vicinity of the Virgo 
 cluster. 

 The influence of magnetic fields on gamma ray propagation is mostly
 restricted to the rate of energy loss in propagation. Fig.~\ref{ts:fig6}
 shows with a dotted line the synchrotron energy loss of electrons
 in 1 nG magnetic field.
 In the absence of magnetic fields electrons would have inverse Compton
 interactions on MBR that would generate a gamma ray that carries
 almost the total electron energy. This would lead to electromagnetic
 cascading and relatively slow decrease of the $\gamma$-ray energy.
 In the presence of fields the electrons of energy above 3$\times$10$^{18}$ eV
 would lose energy very fast
 on synchrotron radiation. The radiated photons would have energies 
 lower by many orders of magnitude. This is one of the bases for
 limiting the allowed mass of the $X$ particles in the exotic 
 particle physics models of the UHECR origin.

\subsection{Production of Secondary Particles in Propagation}

 One interesting feature that can be used for testing of
 the type and distribution of UHECR sources, that we shall not discuss
 at any length, is the production of secondary particles in
 propagation. The energy loss of the primary protons and $\gamma$-rays
 is converted to secondary gamma rays and neutrinos.

 Most interesting are the cosmogenic neutrinos, that were first 
 proposed by Berezinsky \& Zatsepin~\cite{BZ68} and have been since
 calculated many times, most recently in Ref.~\cite{ESS01}.
 Every charged pion produced in a photoproduction interaction
 three neutrinos through its decay chain. 

 The spectrum of cosmogenic neutrinos depends on the UHECR 
 spectrum and on the UHECR source distribution. It extends
 to energies exceeding 10$^{20}$ eV. Currently designed and built 
 neutrino telescopes are aiming at detection of cosmological
 neutrinos (see P.~Gorham's talk).

 At the $\Delta$ resonance energy range 2/3 of the produced pions
 are neutral. Most of the energy loss (including those in $e^+e^-$
 pairs) goes to the electromagnetic component as do the muon
 decay electrons. The ensuing electromagnetic cascading should
 create a $\gamma$-ray halo around powerful UHECR sources that
 could be detected by the new generation of $\gamma$-ray 
 detectors.

\subsection{Particle Physics Models}

  There are also quite a few models that attempt to avoid the
 propagation difficultios by assuming that UHECR are not any
 of the known stable particles. One of these models~\cite{GF96,GFPLB98}
 assumes that the gluino is the lightest supersymmetric particle
 and that it causes the observed UHE cosmic ray showers. It gluinos
 have mass between 0.1 and 1 GeV, it could have photoproduction cross
 section much smaller than that ot nucleons and still be able to 
 interact in the atmosphere and generate showers. The distance 
 to potential gluino sources could reach many hundreds of Mpc.
 The problems with this and other similar models that neutron 
 particles can not be accelerated and gluinos have to be
 copiously produced at the source in interactions of particles 
 of still higher energy. So the problems again lead to the 
 mechanisms of particle acceleration well above 10$^{20}$ eV.

  Starting with Refs.~\cite{GM96,CG97} many authors have discussed
 different effects that could help propagate protons and much 
 larger distances than calculated above. The original suggestions
 are for tiny violations of the Lorentz invariance that can not be
 detected in any other way. Coleman \& Glashow define a maximum 
 achievable velocity (MAV) for the particles involved in photoproduction
 interactions that are due to Lorentz invariance violations. 
 Small (order of 10$^{-23}$ differences between proton and $\Delta^+$
 or proton and pion MAV would increase the photoproduction
 threshold and thus significantly increase the energy of the GZK 
 cutoff. 

\section{EXPERIMENTAL DATA}

  There are six experiments that have collected experimental data 
 after the Volcano Ranch array that Linsley ran. SUGAR, Haverah Park,
 Yakutsk, and AGASA are surface air shower arrays. Fly's Eye and its
 successor HiRes are fluorescent detectors. The current world 
 statistics on UHECR are dominated by AGASA and HiRes. 
 These two experiments have presented data on the energy spectrum,
 some shower features ($X_{max}$) and arrival direction distributions
 for their data samples. The statistics is still quite small and the
 main results of the two detectors are somewhat in contradiction.

\subsection{Energy Spectra of UHECR}

\begin{figure}[thb]
\includegraphics[width=85mm]{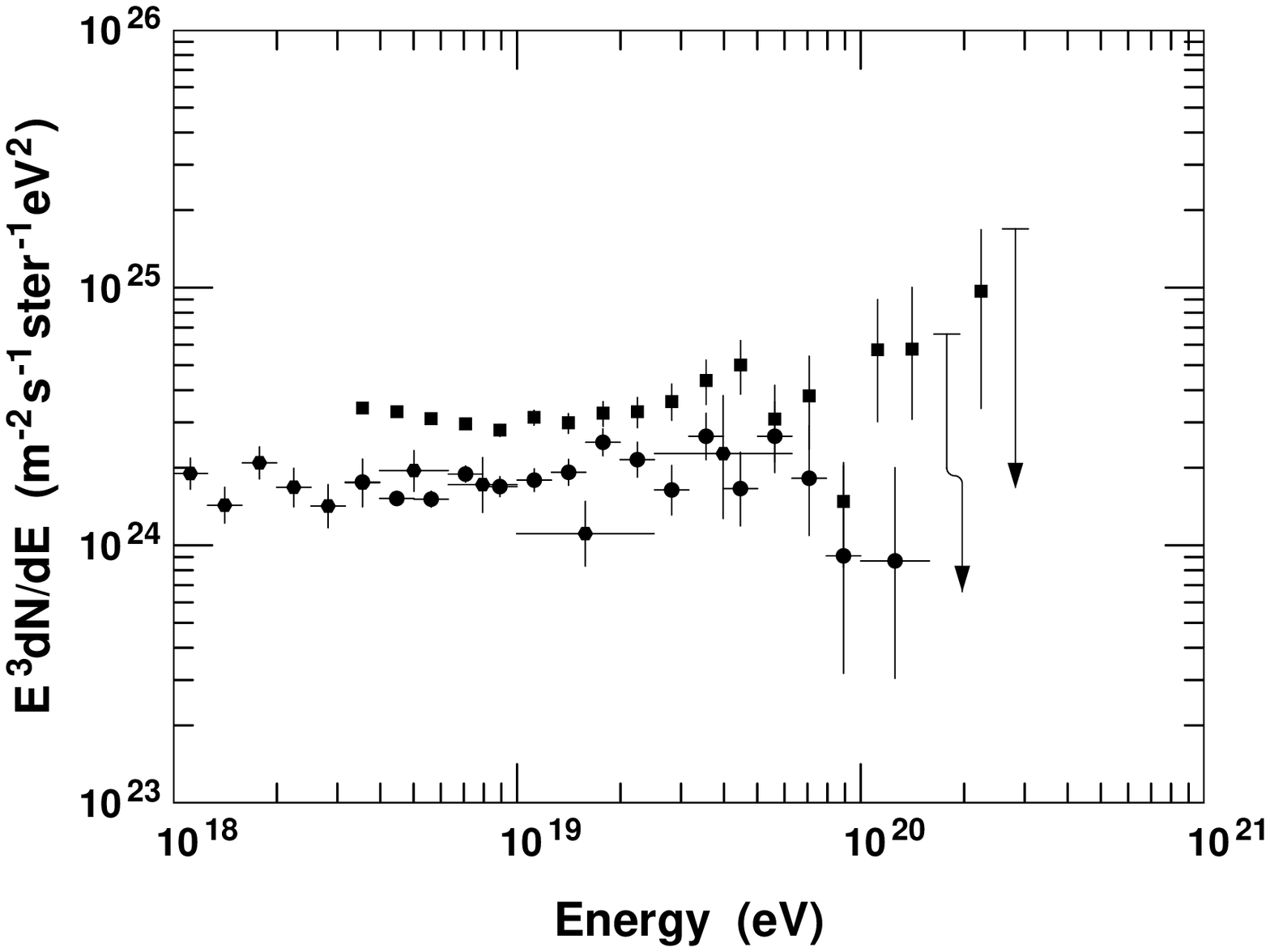}
\includegraphics[width=85mm]{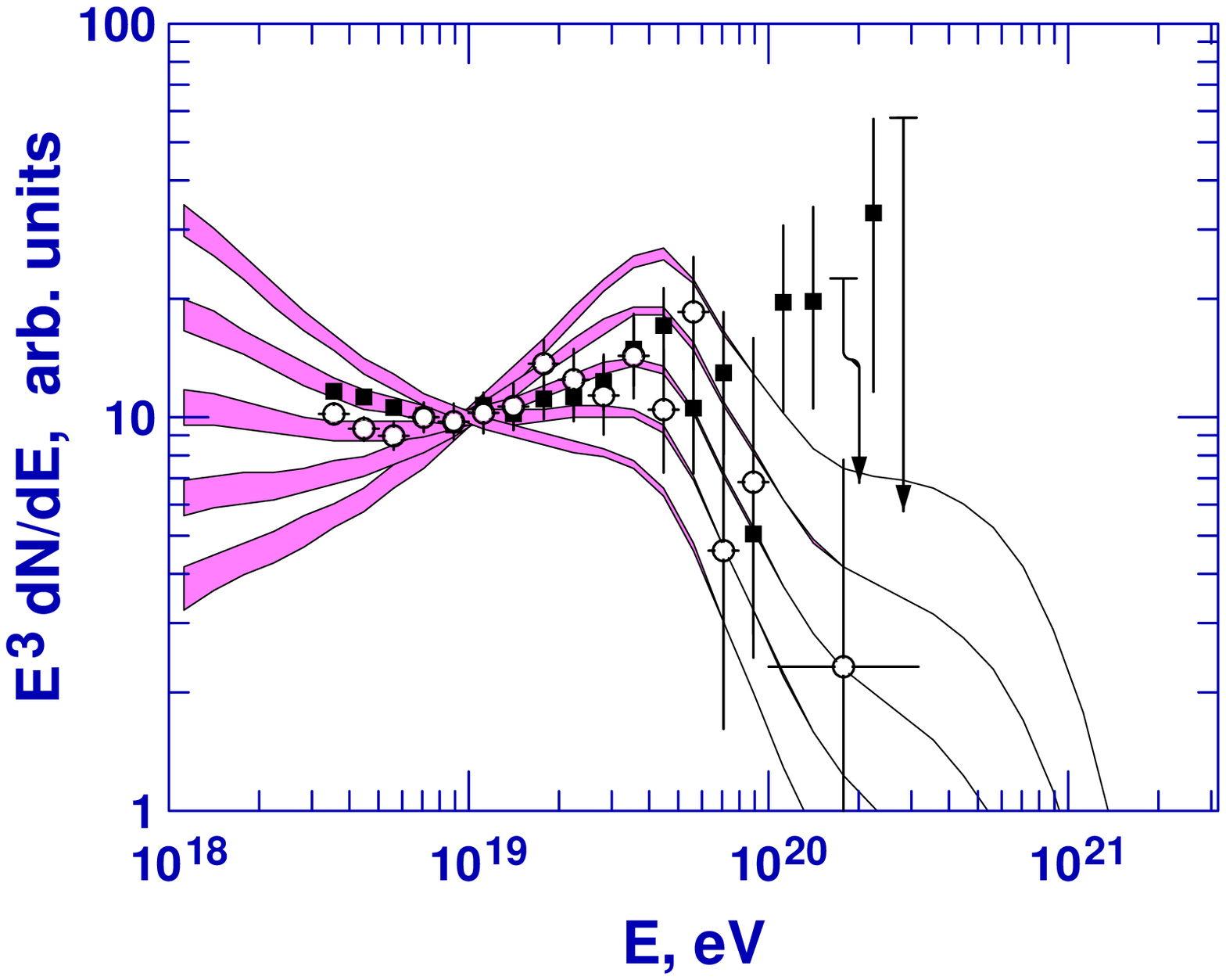}
\caption{ Left hand panel: the experimental data of AGASA (full squares)
 and of HiRes in monocular mode (dots).
 Right hand panel: The data are normalized to each other at 10\protect$^{19}$
 eV and are compared in shape to calculations such as shown in
 Fig.~\protect\ref{ts:fig8} with injection spectra with indices
 of 2.00, 2.25, 2.50, 2.75, and 3.00 (bottom to top at
 10\protect$^{18}$ eV.
}
\label{ts:fig9}
\end{figure}

 Figure~\ref{ts:fig9} shows in the left hand panel the UHECR spectra
 measured by  AGASA~\cite{AGASA} and by HiRes~\cite{HiRes1,HiRes2}
 in monocular mode. This experiment is designed 
 to look for showers {\em in stereo}, with two optical detectors
 that observe the shower simultaneously. The shower analysis is much
 easier and exact in such a mode. The data shown in Fig.~\ref{ts:fig9}
 are taken  by the two optical detectors independently. The 
 experimental statistics of HiRes in stereo is still smaller.

 There two facts that one immediately notices:\\
 1) The overall normalization of the cosmic ray flux is different
 by about 40\%, which looks like a factor of 2 in the figure because
 of the $E^3$ factor on the differential flux.\\
 2) AGASA sees eleven showers with energy above 10$^{20}$ eV. HiRes sees 
 only two with a an exposure that is estimated twice the AGASA's.

 The energy assignment of the AGASA experiment is done by the particle
 density at 600 meters from the core ($\rho_{600}$) which is not very
 sensitive to the type of the primary nucleus. HiRes estimates
 the primary energy either by the integral of the shower profile.
 Both experiments claim systematic uncertainties of order 30\%.
    
 The data from the other experiments are intermediate between these
 two extremes. Each experiment has seen at least one event that has 
 energy well above 10$^{20}$ eV, but with small experimental 
 statistics the flux of such events is very uncertain.

 When the difference in flux normalization (which could well be a 
 difference in energy assignment) is taken out, the spectral shapes
 are not that different, as shown in the right hand panel of 
 Fig.~\ref{ts:fig9}. The spectra agree in shape quite well up
 to 10$^{20}$ eV. They may show a dip at about 10$^{19.7}$ eV.
 The big difference is that the AGASA spectrum recovers and 
 that of HiRes does not. Because of the small statistics the
 difference between the two data sets at the highest energies
 is not statistically significant~\cite{BdMO}.

 Another question that Fig.~\ref{ts:fig9} raises is on the 
 UHECR injection spectrum. The eyes (and numerous fits) prefer
 a spectral index in the vicinity of 2.50 in the case of
 homogeneous isotropic source distribution. Flatter injection 
 spectra could be fitted only by a strong contribution of 
 the galactic cosmic rays up to 10$^{19}$ eV. Distinguishing 
 between galactic and extragalactic contribution would only
 be possible if the chemical composition of UHECR is known - 
 galactic cosmic rays at this energy have to be heavy nuclei.

 Another difference between the two measurements, which is not
 that easy to see, is that they observe the ankle at different
 energy: about 3$\times$10$^{18}$ eV by HiRes and about
 10$^{19}$ eV by AGASA. This is obviously not related
 to the energy assignments discussed above, which only differ
 by 40\% or so.  
 
\subsection{Chemical Composition of UHECR}
  
 The predecessor of HiRes, the Fly's Eye, studied the UHECR composition
 by fitting the energy dependence of the depth of maximum $X_{max}$ 
 of the UHE air showers. In electromagnetic showers $X_{max}$ is 
 proportional to logarithm of the primary energy $\ln E$. In showers
 initiated by primary nuclei this dependence is a more complicated
 function of the energy, but is still proportional to the primary
 energy per nucleon. Because of that $X_{max}$ of an iron initiated
 shower of energy $E_0$ would be approximately the same as of a proton
 shower of energy $E_0$/56. 
 
 The change of $X_{max}$ in one decade of $E_0$ was called elongation
 rate (ER) by John Linsley. Contemporary hadronic interaction models
 predict elongation rate of 50 to 60 g/cm$^2$. The Fly's Eye observed
 an ER of more than 70 g/cm$^2$ to energies 
 3$\times$10$^{17}$ eV followed by a change that made the elongation
 rate consistent with expectations~\cite{TKGetal93}. The Fly's Eye thus
 observed a simultaneous change of the slope of the cosmic ray spectrum 
 and of the shower elongation rate~\cite{Birdetal93}. 

\begin{figure}[thb]
\includegraphics[width=95mm]{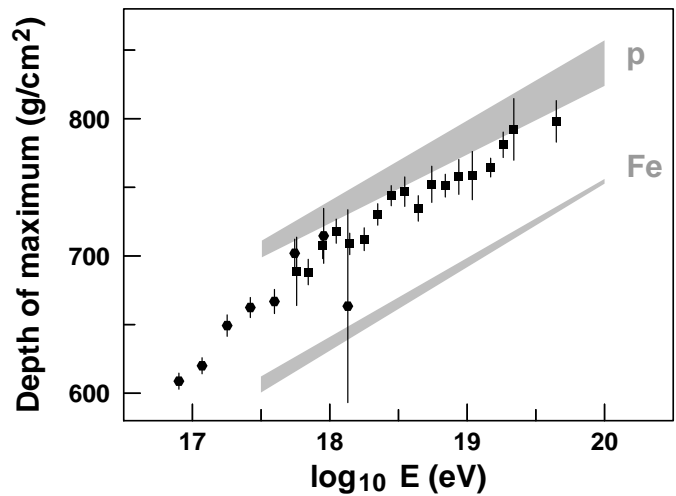}
\caption{Depth of maximum as a function of the shower energy, as
 measured by HiRes (squares) and HiRes prototype + CASA/MIA 
 (hexagons). The shaded areas are the predictions of different
 hadronic interaction models.
 }
\label{ts:fig10}
\end{figure}
 
 This is an enormously important statement, because it implies that
 the the mass of the primary cosmic rays is becoming lighter. 
 It signals that the flux of heavy galactic cosmic rays is decreasing
 and the new cosmic ray component that is responsible for the change
 of the spectral index at the ankle consists of protons and possibly
 He nuclei. 

 This observation is now supported~\cite{HRcomp}
 by the new results of HiRes shown in Fig.~\ref{ts:fig10}.
 The lower energy points shown with hexagons are obtained 
 with the HiRes prototype working together with CASA/MIA
 air shower array. According to this most recent data set
 the change of the cosmic ray composition is almost over
 by 10$^{18}$ eV. Since the observed X$_{max}$ is almost 
 parallel to the expectations, one can assume that the 
 composition is constant. The composition is proton dominated,
 but a more exact prediction depends on the hadronic interaction
 model used, as shown with shaded areas that indicate calculations
 with different hadronic models.

 The chemical composition of the primary cosmic rays can also be studied
 by the fraction of muons in the showers. When the primary particle
 is a heavy nucleus, the first generation mesons are approximately
 1/A less energetic than in protons showers, and have a higher decay
 probability. For this reason the muon fractions is higher in showers
 generated by heavy nuclei. The AGASA experiment has a number of muon
 counters that can study the muon content of the showers. 
 The number of such counters is not high and the statistics is thus
 limited, but the AGASA collaboration
 did not see the change in the muon content that would correspond
 the the Fly's Eye result. The collaboration still claims a gradual
 decrease of the Fe fraction between energies of 10$^{17.5}$ and
 10$^{19}$ eV. Alternative methods for estimation of the cosmic
 ray composition, applied by other experimental groups, also 
 give a relatively large fraction of heavy nuclei around 10$^{18}$ eV. 

 This disagreement is still not fully solved. The showers of both groups are
 now analyzed using the same hadronic interaction models but the 
 differences persist.

 Especially important is the determination of the type of the UHECR - 
 nuclei or gamma rays. Two special studies have been performed
 by the AGASA and the Haverah Park groups using different approaches.
 AGASA~\cite{agasa_g}
 looks at the particle density at 100 meters from the shower core that
 is expected to be dominated by shower muons. This density should be
 much lower in gamma ray initiated showers. A new analysis of the Haverah
 Park data~\cite{hp_new}
 studies highly inclined air showers. Since the absorption of gamma ray
 showers is stronger, their flux should decline with zenith angle faster.
 Both experiments limit the fraction of $\gamma$-rays above 10$^{19}$ eV
 at 30\% of all cosmic rays. The limits at higher energy
 (3-4$\times$10$^{19}$ eV) are less strict(67\% - 55\%)  because of
 the declining statistics. There is no statement about the few particles
 above 10$^{20}$ eV so that `top-down' models can still apply to these
 events. 
 
 \subsection{Arrival Directions of UHECR}

\begin{figure}[thb]
\includegraphics[height=65mm]{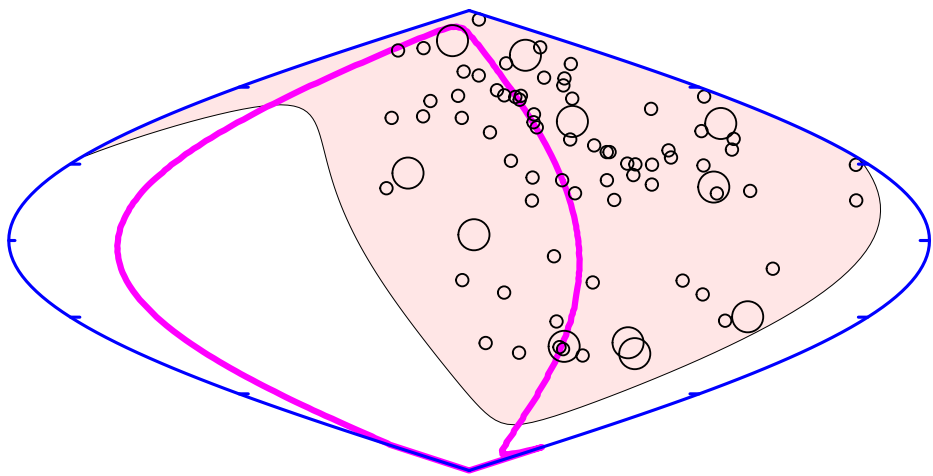}
\caption{Arrival directions of cosmic rays of energy exceeding
 10\protect$^{19.6}$ eV in galactic coordinates. The map is centered
 on the galactic anti-center. The large circles identify energies exceeding
 10\protect$^{20}$ eV. The shaded area is the field of view of the 
 detectors. The thick line indicates the supergalactic plane - the
 plane of weight of galaxies at redshifts less than 0.04.
}
\label{ts:fig11}
\end{figure}

 The arrival direction distribution of the UHECR events detected by AGASA
 has two main features: the distribution is isotropic on large scale and
 non isotropic on small (one degree) scale~\cite{agasadir}.
 There is no preference of higher event rate coming from the galactic
 plane or any other know astrophysical concentration of matter,
 although an association with the supergallactic plane was
 reported~\cite{tsetal95} on the basis of an earlier smaller event sample.
 There are however five doublets and a triplet of events arriving at less
 than 2.5$^\circ$ from each other. The angular resolution of the detector  
 is below 2$^\circ$.  The statistical significance of 
 this `clustering' is of order three $\sigma$. The individual clusters 
 can be identified in Fig.~\ref{ts:fig11} that shows the arrival directions
 of the world data set above 10$^{19.6}$ eV, except for the Fly's Eye and
 HiRes events that are not yet published. 

 The clusters do not point at any known luminous astrophysical object.
 There have been long discussions how such clustering can occur in
 a more or less realistic astrophysical scenario. Do they point at 
 UHECR sources that we cannot otherwise observe? How many UHECR sources
 should exist to fit both the isotropic large scale distribution and
 the clustering? Is the clustering real, or a result of a large 
 statistical fluctuation? 

\begin{figure}[thb]
\includegraphics[height=60mm]{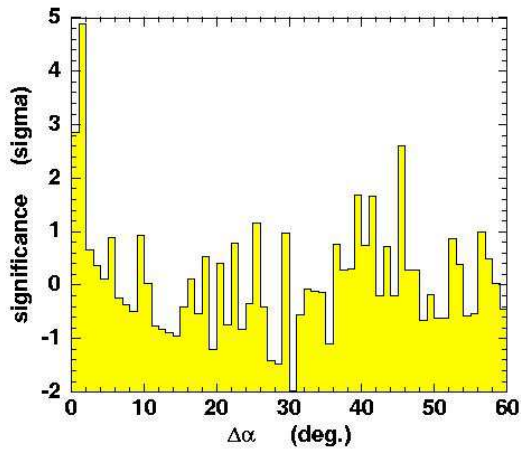}
\caption{ Self correlation plot of the arrival directions of the
AGASA showers of energy above \protect4$\times$10$^{19}$ eV.
}
\label{ts:fig12}
\end{figure}

 One of the better ways to look at the small scale anisotropy of 
 the AGASA showers is to look at the self-correlation of the 
 detected showers. Fig.~\ref{ts:fig12} shows the self-correlation
 plot - the number of events that have directions different from each
 other by $\Delta \alpha$. The real number of events is divided 
 by the expectation from isotropic distribution and is given
 in the figure in terms of $\sigma$. The clustering is at 
 $\Delta \alpha$ less than 2$^\circ$, which would correspond 
 to the angular resolution of the detector. The statistical 
 significance reaches 5$\sigma$, significantly higher than 
 simply counting the number of clusters.

 Unfortunately there is again a controversy between the observations
 with the highest statistics - the HiRes collaboration does not
 confirm~\cite{HR_clust} the clustering of events reported by AGASA.
 The much more accurate stereoscopic event reconstruction does
 not find~\cite{HR_anis} small scale anisotropy. It is difficult to
 compare directly the results of the two groups because the HiRes 
 event sample above 10$^{19}$ eV is smaller. HiRes, however,
 claims that there is no statistically significant anisotropies on
 angular scales up tp 5$^\circ$ at any energy threshold above
 10$^{19}$ eV.
  
 A detailed review of the experimental approaches and data before
 the most recent HiRes publications is this of Nagano \$ Watson~\cite{N&W}
 The main qualitative and contradictory experimental results obtained
 by the  current experiments with the highest statistics are summarized
 below.\\[1truemm]

 \begin{tabular}{r c c}
  &  AGASA  &  HiRes  \\
 are there super GZK events & many & a few \\
 have we observed GZK cutoff & yes & no \\
 is there large scale isotropy & yes & yes \\
 is there small scale isotropy       & no & yes \\
 UHECR composition changing at& 3\protect$\times$10$^{18}$ eV & 
 10\protect$^{18}$ eV \\
 \end{tabular}
\vspace*{2truemm}

  We were nevertheless able to estimate, although very roughly, 
 the main astrophysical parameters related to ultra high energy
 cosmic rays. The emissivity of particles of energy between 
 10$^{19}$ and 10$^{21}$ derived for $\alpha$ = 2 injection
 spectra~\cite{Wax95} is 4.5$\times$10$^{44}$ ergs/Mpc$^3$/yr.
 For injection spectra of index 2.7 the emissivity in the same
 units is above 10$^{46}$~\cite{BGG02}. There is a strong
 dependence of the emissivity at all energies on the steepness of
 the injection spectrum and on the {\em minimum} of the energy 
 spectrum at acceleration. 

  The density of the UHECR sources has been estimated from the 
 clustering of the AGASA cosmic rays events~\cite{BdM04,SME04,YNS03}
 to 10$^{-{5 \pm 1}}$ Mpc$^{-3}$
 depending on the assumptions for the extragalactic magnetic fields.

  The energy spectrum detected by AGASA can be explained by a combination
 of isotropic homogeneous UHECR source distribution, galactic cosmic 
 rays and local UHECR sources or top-down scenarios. The HiRes 
 spectrum does not require local sources or top-down models.

  We have yet no idea what the cosmological evolution of the UHECR 
 sources is. Steep injection spectra do not require any cosmological
 evolution. Flat injection spectra do.

\section{FUTURE OF THE FIELD}

 It is obvious that at least a part of the inconsistency of the 
 current experimental results is due to the very low experimental
 statistics. The interest in UHECR actually grew when ideas for
 experiments that can solve the problem appeared in the early 1990s.
 Obviously shower arrays  much bigger than AGASA are needed when
 the aim is to collect reasonable statistics above 10$^{20}$ eV.
 The first idea for the Auger Observatory~\cite{Auger} was 
 spelled out by Cronin \& Watson. 

 Auger would initially consist of two air shower arrays located in 
 the Northern and in the Southern hemispheres, each of area 
 5,000 km$^2$, i.e. it was supposed to be 100 times the size 
 of AGASA. Arrays in both hemispheres are needed for full sky 
 coverage - nobody knows where the UHECR sources are. This first
 proposal is scaled down, but the Southern Auger Observatory 
 is being built in Argentina (see talk of R.~Cintra Shellard).
 It consists of 3,000 km$^2$ area
 observed by surface detectors and by four fluorescent detectors.
 The surface detectors are 10 m$^2$ water Cherenkov tanks spaced 
 by 1,500 m. One meter tall water Cherenkov tanks were chosen 
 because they absorb almost fully all shower electrons and photons
 and because their area does not decrease that fast with zenith
 angle - they will be able to observe and reconstruct showers
 down to horizontal direction. Another big advantage is that about
 10\% of the statistics will be observed simultaneously with both
 techniques, which may solve the current contradictions of AGASA 
 and HiRes. 

 Then there is the Telescope Array (TA)~\cite{TA} that is already
 funded by the  Japanese government. TA will consist of 1,000 km$^2$
 surface array equipped with scintillator detectors 1.2 km apart from
 each other and three fluorescent detectors observing the atmosphere
 above it. There are plans for moving the HiRes detectors to TA
 and infilling the ground array to extend the threshold to lower 
 energy. The latter part is not yet funded.

 Other, longer time scale, plans are for space based giant air shower
 detectors. EUSO~\cite{EUSO} was chosen by the European Space Agency
 as a Phase A project to be mounted on the International Space Station.
 EUSO is a fluorescence detector that observes the atmosphere from 
 an altitude of 400 km. It will be able to cover more than 10$^5$ km$^2$
 with a resolution about 1 km per pixel. Such a detector will be very
 efficient for observation of highly inclined showers, and thus it would
 be a very good detector of UHE neutrinos that would interact very 
 deep in the atmosphere. EUSO was not approved for Phase B because
 of  the uncertainty of the ISS future, but the design of the detector
 and the necessary measurements of the relevant atmospheric features 
 are still funded and continue.

 OWL~\cite{OWL} is an even more ambitious project for stereoscopic
 observations from space with two detectors mounted on free flying 
 satellites. These satellites would see the Earth from higher altitude,
 and thus observe larger area. The stereoscopic observation will make
 the analysis of the detected events easier and more exact.
 OWL is a part of the NASA plans for the future.

\section{SUMMARY}

 We are certain that cosmic rays of energy above 10$^{20}$ eV exist,
 but their flux is unknown. All giant air showers arrays have 
 detected at least one shower of energy above 10$^{20}$ eV. The puzzle 
 is that very few astrophysical objects can accelerate charged nuclei
 to such energy in shock acceleration processes, which is the favorite 
 model for astrophysical acceleration.

 Protons and heavier nuclei lose energy in propagation in photoproduction
 and photodisintegration on MBR and other photon fields. Since the 
 energy loss length of all nuclei is of order 10 Mpc, the UHECR sources
 have to be within tens of Mpc from our Galaxy. There are few
 powerful astrophysical objects that close to us.

 The other possibility are `top-down' scenarios where these particles
 are generated in the decay of ultraheavy X-particles, which could be
 emitted by cosmic strings or are long lived remnants of the early
 Universe.

 The current experimental data are not able to give us good indication
 on the type of these UHECR and their arrival direction distributions.
 The data on the cosmic ray spectrum and composition above 10$^{18}$ eV
 are somewhat contradictory. The HiRes experiment gives UHECR spectrum
 and composition fully consistent with the assumption of isotropic
 distribution of the cosmic rays sources. The HiRes predecessor, Fly's Eye,
 has detected the highest energy event of energy 3$\times$10$^{20}$ eV.
 The AGASA experiment has published several events of super GZK energy 
 which show no cutoff in the UHE cosmic ray spectrum. AGASA also 
 claims a small scale anisotropy of the events above 10$^{19}$ eV
 which is not, however, related to the directions of powerful 
 astrophysical objects. 

 New third (and fourth) giant air shower experiments are being designed
 and built. They will increase the world data sample by orders of 
 magnitude and help understanding the nature and sources of these
 exceptional events. 

 {\bf Acknowledgments} My work in the field of UHECR is funded in part
 by U.S. Department of energy contract  DE-FG02 91ER 40626 
 and by NASA grant NAG5-7009. The collaboration of J.~Alvarez-Mu\~{n}iz,
 P.L.Biermann, R.~Engel, T.K.~Gaisser, D.~Seckel and others is highly
 appreciated. 


\end{document}